\documentclass[amsmath,amssymb,
aps,
nofootinbib,
prd,
superscriptaddress,
tightenlines,
notitlepage,
twocolumn,
showpacs,
floatfix]{revtex4-2}
\usepackage[utf8]{inputenc}
\usepackage{color}
\usepackage{xcolor}
\usepackage{hyperref}
\usepackage{graphicx}
\usepackage{dcolumn}
\usepackage{bm}

\newcommand{\DOA}{\affiliation{Department of Astronomy, School of Physics,
Peking University, Beijing 100871, China} }
\newcommand{\KIAA}{\affiliation{Kavli Institute for Astronomy and
Astrophysics, Peking University, Beijing 100871, China}}
\newcommand{\NAOC}{\affiliation{National Astronomical Observatories, 
Chinese Academy of Sciences, Beijing 100012, China} }

\begin{document}

\preprint{APS/123-QED}

\title{Dark radiation as a probe for phase transition in the early universe}
\author{Zihang Wang}\email{wzhax@pku.edu.cn}\DOA\KIAA
\author{Lijing Shao}\KIAA\NAOC
\date{\today}
\begin{abstract}
The cosmological constant is not necessarily small in the early universe. If a scalar field obtains a vacuum expectation value after a phase transition (PT), a possibly large cosmological constant could present before PT. The early cosmological constant (ECC) and the PT process may be detectable from dark radiation (DR) today, such as in the cosmic axion background. We show that for a broad class of DR models, the DR density and spectrum are significantly modified by the presence of an ECC. From the density and the spectrum of the DR today, we can deduce the temperature and the strength of the PT.
\end{abstract}

\maketitle

\emph{Introduction.---}The cosmological constant is an enigma in theoretical physics~\cite{CC1a,CC1b,CC1}. Its value observed today is very close to zero~\cite{CC3}. However, in quantum field theory, an unacceptably large cosmological constant $\Lambda_{\rm qu}$ is expected from vacuum fluctuations. So an incredibly precise cancellation must be present between the bare cosmological constant $\Lambda_{\rm B}$ and $\Lambda_{\rm qu}$~\cite{CC1}. Various scenarios have been proposed~\cite{CC4,CC5}, but the cosmological constant problem is unsolved yet.

Even if we manage to achieve cancellation and obtain a small enough cosmological constant today, a possibly large cosmological constant could be present in the early universe~\cite{CC2}, which naturally comes from phase transitions (PTs). In many extensions of the Standard Model, a first-order PT occurs in the early universe~\cite{GW3}. PTs occur when a scalar field, such as the Higgs field---or some combination of fields---obtains a non-zero vacuum expectation value (VEV). Such a process leads to a change in vacuum energy~\cite{ECC1}. If we require the cosmological constant almost vanish today, it has to be large before the PT.

Early dark energy was widely studied recently because they may help solve the Hubble tension~\cite{EDE,EDE2,EDE3}. Early cosmological constant (ECC) may also play an important role before big bang nucleosynthesis (BBN). It can be induced by a scalar field and leads to cosmic inflation~\cite{inflation1,inflation2}. ECC can also appear before a PT. Recently a lot of studies focused on production of gravitational waves from early-universe PTs~\cite{GW1,GW2,GW3,GW4}. A strong first-order PT can produce detectable gravitational waves~\cite{gwpt1,gwpt2,gwpt3,gwpt4}. If ECC is important before a PT, the standard cosmology is modified. It is possible that the universe is not radiation dominated before BBN. Such non-standard cosmologies were considered, mostly focused on early matter domination~\cite{NSC3}, early kination domination~\cite{NSC4} and low temperature reheating~\cite{NSC5}.~\citet{ax1} showed these non-standard cosmologies to affect axionic dark matter density today. \citet{NSC2} considered thermal axion densities in non-standard cosmologies, including a low temperature reheating scenario and a kination scenario. It has also been proposed to probe the early universe from axionic dark matter today~\cite{ax2}.

Dark radiation (DR) has the potential to become another messenger of the early universe in addition to gravitational waves. Among the candidates of DR, a particular interesting example is axions or axion-like particles (ALPs). These particles are physically well motivated and interact weakly with the Standard Model~\cite{ax3,ax4}. They are created in the early universe, and may contribute to both dark matter and DR in different axion production mechanisms~\cite{marsh}. Such cosmic axion background (CAB) could be detectable, using ADMX-like resonant cavity~\cite{CaB}. Other DR candidates include dark photons~\cite{DP1,DP2}, minicharged particles~\cite{DP2,MCP1}, {\it etc.}. In this paper, we study the effects of ECC and PTs on DR density and spectrum. We will show that it is possible to extract the detailed evolution of early universe from today's DR density and spectrum.

\emph{Phase transition.---}Consider a PT induced by a scalar field. At high temperatures the scalar field is in symmetric phase, and we take its VEV as $\langle\phi\rangle=0$. As the temperature of the universe dropped to $T_{c}$, the critical temperature of the PT, a second minimum at $\phi=\phi_{c}$ of the effective potential of the scalar field
becomes degenerate with the minimum of $\phi=0$~\cite{GW1}. When $T<T_{c}$, bubbles start to nucleate from the false vacuum. These bubbles expand and collide until all of the false vacuum disappears. The percolation temperature~\cite{GW1,percolation} $T_{p}$ is the temperature that the probability of false vacuum is roughly $71\%$. The temperatures $T_{c}$ and $T_{p}$ correspond to cosmic time $t_{c}$ and $t_{p}$, respectively.

We assume that the PT occurred after inflation. To avoid effects on BBN, we require $T_{p}\gg 1\,{\rm MeV}$. The ratio between the ECC density and the radiation density at $T_{p}$ is $\alpha_{\Lambda}=\rho_{\Lambda}(T_{p})/\rho_{R}(T_{p})$, representing the strength of PT. $\alpha_{\Lambda}$ can naturally reach order unity for a strong PT~\cite{GW1}. The ECC decay rate when PT occurs is $\Gamma_{\Lambda}$. The decay rate per volume is $\Gamma_{\Lambda}/V\sim A(T)\exp\left[-S_{3}(T)/T\right]$,
where $S_{3}(T)$ is the three dimensional Euclidean action~\cite{GW1}. For early universe PTs, thermal fluctuations dominate the decay rate. In such a case, the prefactor is $A(T)\sim T^{4}[S_{3}(T)/(2\pi T)]^{3/2}$, which is independent of initial bubble radius.

The comoving bubble radius at $t'$ for a bubble nucleated at time $t$ is~\cite{GW1},
\begin{equation}\label{eq:radius}
r(t',t)=\int_{t'}^{t}\,\mathrm{d}t'' v_{w}(t'')\frac{a(t)}{a(t'')}   \, , 
\end{equation}
where $v_{w}(t'')$ is the bubble wall velocity, $a(t)$ is the scale factor. The false vacuum probability is $P(t)=e^{-I(t)}$~\cite{GW1}, where $I(t)$ is a weight function,
\begin{equation}\label{eq:we}
I(t)=\int_{t_{c}}^{t}\,\mathrm{d}t' \Gamma(t') \left(\frac{a(t')}{a(t)}\right)^{3} \left[\frac{4\pi}{3}r(t',t)^3\right]   \,.
\end{equation}

For fast PTs, We can neglect the change of scale factors during PT and $v_{w}$ can be taken as a constant. We can also expand $\Gamma(t)$ near the percolation time~\cite{GW1}, $\Gamma(t)=\Gamma(t_{p})e^{\beta(t-t_{p})}$,
where the factor $\beta$ characterizes the duration of PT. A dimensionless parameter $\tilde{\beta}$ is defined as $\tilde{\beta}=\beta/H(t_{p})$. For fast PTs
we can integrate Eq.~(\ref{eq:we}) to obtain $I(t)=I_{0}e^{\beta(t-t_{p})}$, where $I_{0}=8\pi v_{w}^{3}\Gamma(t_{p})\beta^{-4}$. Using $P(t_{p})=0.71$ we find $I_{0}=0.34$.



\emph{Phase transition effects on DR.---}We consider an ECC with an initial density $\rho_{\Lambda 0}$. During a PT, the false vacuum decays and the energy converts to kinetic energy of fluid and sound waves propagating through it (we neglect the energy of sound waves here). Then the motion of fluid dissipates to reheat the universe. If we assume that the universe is still uniform on the scale $L\gg R(t)$, where $R(t)$ is the mean bubble separation, we can average these velocities as $\langle v_{i}v_{j}\rangle=v^{2}\delta_{ij}/3$. We find that on average the contribution of kinetic energy to cosmic expansion is the same as radiation, with an equation of state $p=\rho/3$. Hence after averaging over a length scale $L$ we find the total energy density except ECC, $\rho_{t}=\rho_{R}+\rho_{\rm kin}$, behaves the same way as radiation. So we have
\begin{equation}
  \frac{1}{a^{4}}\frac{\mathrm{d}}{\mathrm{d}
  t}\left(\rho_{t}a^{4}\right)=\Gamma_{\Lambda}\rho_{\Lambda 0}P(t)\, .
\end{equation} 

After the PT we take the effective particle numbers, $g_{*}(T)$, and the effective entropy degrees of freedom, $g_{*S}(T)$, from Ref.~\cite{grgs}. Before and during the PT, the temperature is of order ${\cal O}(100\,{\rm GeV})$. Hence it is a good approximation to treat $g_{*}(T)$ and $g_{*S}(T)$ as constants.


In the following we will consider three classes of different DR production mechanisms. 
\begin{itemize}
	\item {\bf Model I}: A heavy non-relativistic parent particle X spontaneously decays into two particles, Y and Z, where Y is a stable light particle and becomes DR today, while Z may be identical to Y (Model Ia) or a particle much heavier than Y (Model Ib).
	\item {\bf Model II}: A non-relativistic particle pair of $\rm X'$ annihilate into two light stable particles $\rm Y'$ which become the DR today. 
	\item {\bf Model III}: DR is created with a temperature-independent intrinsic spectrum that can not be approximately treated as a $\delta$-function like in Models I and II, but has a width. For example, the heavy non-relativistic particle in Model I may decay to three particles, one of which becomes DR. Another possibility is the Doppler broadening of Model I caused by the thermal velocity of X.  
\end{itemize}
In all models discussed here, axions can play the role of DR. Axions produced from parent particle decay like in Model I have been discussed in Refs.~\cite{marsh,CaB}.
In this paper, the term ``intrinsic spectrum'' means the DR energy spectrum calculated from DR production mechanisms under a certain temperature, which is independent of cosmology.

We define the energy spectrum as $\rho(\omega)=\rm{d}\rho_{\rm DR} /\rm{d}\omega$, where $\rho_{\rm DR}$ is the DR density. We also define~\cite{CaB} $\Omega_{\rm DR}(\omega)=(1/\rho_{c})\rm {d}\rho_{\rm DR}/\rm{d}\ln\omega$,
where $\rho_{c}$ is the critical density at present.
We neglect the DR particle mass. The energy spectrum can be obtained from,
\begin{equation}\label{eq:DRcreation}
\frac{1}{a^{3}(t)}\frac{\rm{d}}{\rm{d}t}\left[\rho\left(\omega\frac{a(t_{0})}{a(t)}\right)a^{3}(t)\right]=\Gamma\left(\omega\frac{a(t_{0})}{a(t)},\rho_{X},T\right)  \, , 
\end{equation}
where $t_{0}$ is the present time, $\omega$ is the DR energy at present, $\rho_{X}$ represents a set of particle densities from which DR is created. The term on the right-hand side is a quantity related to the DR production rate.

Let us consider Model I first. We neglect the velocity of X at first and come to it later. The densities of X and Y particles satisfy,
\begin{equation}\label{eq:X}
\frac{1}{a^{3}}\frac{\rm{d}}{\rm{d}t}\left(\rho_{X}a^{3}\right)=-\Gamma_{X}\rho_{X} \, ,
\end{equation}
\begin{equation}\label{eq:Y}
\frac{1}{a^{4}}\frac{\rm{d}}{\rm{d}t}\left(\rho_{Y}a^{4}\right)=
\left\{
\begin{aligned}
&\Gamma_{X}\rho_{X}\, , & (\rm {Model\;Ia})\, ,\\
&\frac{m_{X}^{2}-m_{Z}^2}{2m_{X}^2}\Gamma_{X}\rho_{X}\, , & (\rm {Model\;Ib})\, ,
\end{aligned}
\right. 
\end{equation}
where $\Gamma_{X}$ is the decay rate of X, and $m_{X}$ and $m_{Z}$ are masses of X and Z, respectively. The spectrum of DR particles when they are created is very close to a $\delta$-function, 
\begin{equation}\label{eq:gamma}
\Gamma=\left\{
\begin{aligned}
&\Gamma_{X}\rho_{X}(t)\, \delta\left(\omega\frac{a(t_{0})}{a(t)}-\frac{m_{X}}{2}\right)\, , \\
&\frac{m_{X}^{2}-m_{Z}^2}{2m_{X}^2}\Gamma_{X}\rho_{X}\delta\left[\omega\frac{a(t_{0})}{a(t)}-\frac{1}{2}\left(m_{X}-\frac{m_{Z}^2}{m_{X}}\right)\right]\, ,
\end{aligned}
\right.
\end{equation}
where the upper expression is for Model Ia and the lower one is for Model Ib. The coefficient ahead of the $\delta$-function is determined by integrating over $\omega$ in Eq.~(\ref{eq:DRcreation}) and then comparing it with Eq.~(\ref{eq:Y}). Integrating Eq.~(\ref{eq:DRcreation}) over time, one gets
\begin{equation}\label{eq:x1}
\rho(\omega)=\left\{
\begin{aligned}
&\frac{2\Gamma_{X}\rho_{X}(t_{e})}{H(t_{e})m_{X}}\left(\frac{a(t_{e})}{a(t_{0})}\right)^{3}\, , & (\rm {Model\;Ia})\, , \\
&\frac{\Gamma_{X}\rho_{X}(t_{e})}{H(t_{e})m_{X}}\left(\frac{a(t_{e})}{a(t_{0})}\right)^{3}\, , & (\rm {Model\;Ib})\, ,
\end{aligned}
\right.
\end{equation}
where $t_{e}(\omega)$ is the time $t$ that makes the argument of the $\delta$-function in Eq.~(\ref{eq:gamma}) vanish. Note that for Models Ia and Ib the definitions of $t_{e}$ are different.

\begin{figure}[t]
  \centering
  \includegraphics[width=8.5cm]{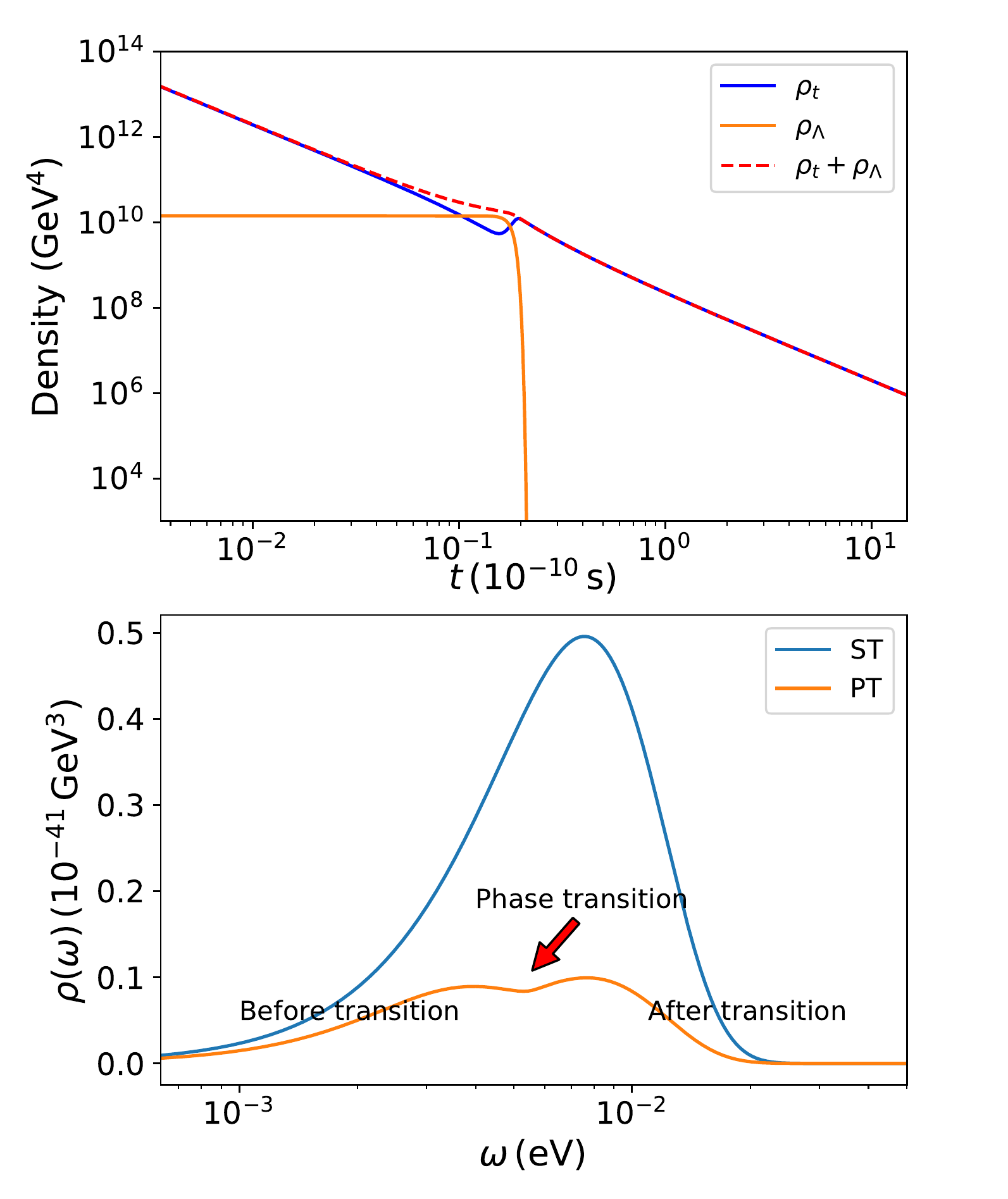}\\
  \caption{(\emph{Top}) Evolution of $\rho_{t}$, $\rho_{\Lambda}$, and total density $\rho_{t}+\rho_{\Lambda}$ in our PT model. (\emph{Bottom}) The DR energy spectrum in Model Ia for standard cosmology (ST) and cosmology with a PT. We have marked the part of the spectrum where DR is produced before, during, and after the PT. We have used $\alpha_{\Lambda}=3$, $T_{p}=100\,{\rm GeV}$, $m_{X}=20\,{\rm TeV}$, and $\Gamma_{X}=3\times 10^{-14}\,{\rm GeV}$.}
  \label{imgd}
\end{figure}

The evolutions of $\rho_{t}$ and $\rho_{\Lambda}$, as well as the effects of the PT on DR spectrum for Model Ia are shown in Fig.~\ref{imgd}. As shown in the bottom panel, PT leaves a kink in DR spectrum $\rho(\omega)$. This is because $\rho_{t}+\rho_{\Lambda}$ changes slope at PT (as can be seen in the upper panel) and $H(t_{e})\propto\sqrt{\rho_{t}+\rho_{\Lambda}}$ in the denominator of Eq.~(\ref{eq:x1}). Although the Hubble parameter is continuous across a fast PT, its time derivative $\mathrm{d}H/\mathrm{d}t$ changes suddenly, which leads to the kink in DR spectrum. We fix $\tilde{\beta}=20$, which is a reasonable value for $\alpha_{\Lambda}\sim 1$~\cite{GW1}. In the figure, each DR energy $\omega$ corresponds to a time in early universe when such DR particles are produced, as marked in the top horizontal axis. For Model I, the DR spectrum can be used to probe PT occurred near $t_{p}\sim 1/\Gamma_{X}$.


Model II can be dealt with similarly. The result is,
\begin{equation}\label{eq:xs}
\rho(\omega)=\frac{\rho_{X'}^{2}(t'_{e})}{m_{X'}^{2}H(t'_{e})}\langle\sigma
v\rangle  \frac{a^{3}(t'_{e})}{a^{3}(t_{0})}\, , 
\end{equation} 
where $\sigma$ is the cross section of pair annihilation, and $v$ is the relative velocity between the two particles. We have considered the case where $\langle\sigma v\rangle$ approaches a constant for low-energy scattering~\cite{W}. Here $t'_{e}$ is the time when the DR particles with a present energy $\omega$ is produced, satisfying $\omega a(t_{0})/a(t'_{e})=m_{X'}$.

\begin{figure}[t]
  \centering
  \includegraphics[width=0.5\textwidth]{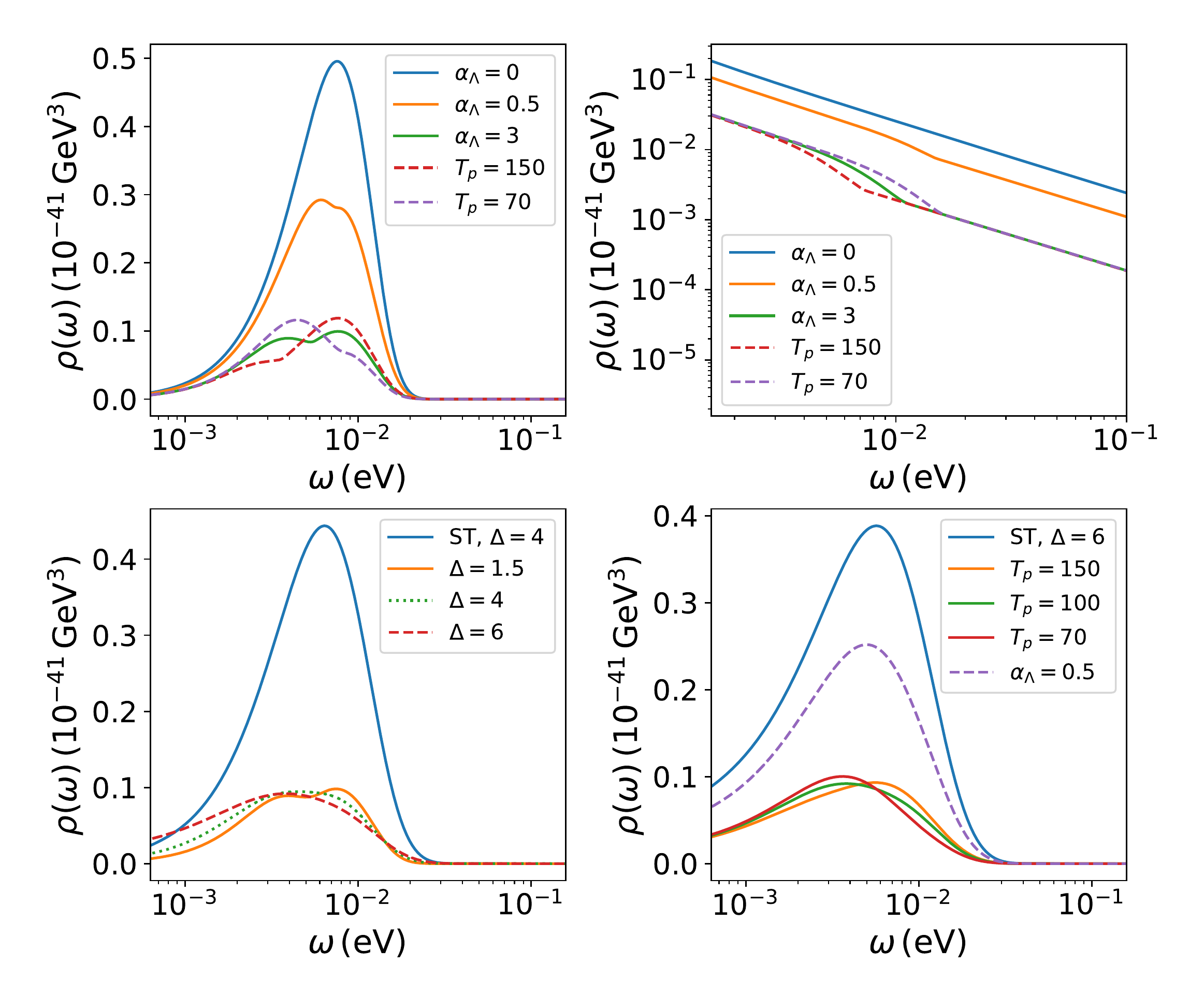}\\
  \caption{DR spectrum for different parameters in Model Ia (\emph{upper left}), Model II (\emph{upper right}) and Model III (\emph{bottom}). For Models Ia and II, $T_{p}=100\,{\rm GeV}$ is fixed and $\alpha_{\Lambda}$ varies for solid lines, while $\alpha_{\Lambda}=3$ is fixed and $T_{p}$ (unit $\rm {GeV}$) varies for dashed lines. For Model III, we keep $\alpha_{\Lambda}=3$ and $T_{p}=100\,{\rm GeV}$ fixed and vary $\Delta$ (unit $\rm {TeV}$) in the bottom left panel. In the bottom right panel $\Delta=6\,{\rm TeV}$ is fixed, and $\alpha_{\Lambda}$ and $T_{p}$ (unit $\rm {GeV}$) vary. $\alpha_{\Lambda}=3$ is fixed for solid lines, while for the dashed line we choose $T_{p}=100\,{\rm GeV}$. We take $m_{X}=m_{X'}=20\,{\rm TeV}$, $\Gamma_{X}=3\times 10^{-14}\,{\rm GeV}$, and $\langle\sigma v\rangle=10^{-48}\,{\rm cm^2}$ here. }
  \label{imga}
\end{figure}

Figure~\ref{imga} shows the DR spectrum when we change the PT strength $\alpha_{\Lambda}$ and temperature $T_{p}$ with $\tilde{\beta}=20$ being kept fixed. For both Models I and II, the upper panel of Fig.~\ref{imga} shows that an ECC followed by a PT leaves a kink in the DR spectrum. The position of the kink of the curve moves leftward as $T_{p}$ increases. For larger values of $\alpha_{\Lambda}$, the kink becomes more apparent. 

The total density of DR decreases if a PT occurred as shown in Fig.~\ref{imga}. The suppression of DR density is mainly due to reheating, during which the density of particles is unchanged while the temperature rises. Hence the universe must expand more to cool down to present cosmic microwave background (CMB) temperature, which dilutes the DR density. In Models I and II, the DR particles created before and after the PT are both diluted. For a fast PT, the temperature rises roughly by a factor $(1+\alpha_{\Lambda}/0.71)^{1/4}$ compared with the standard cosmology. Hence we have approximately $\rho_{\rm DR}/\rho_{\rm DR,s}\approx(1+\alpha_{\Lambda}/0.71)^{-1}$.

For Models I and II, each frequency in DR spectrum corresponds to a specific time in the early universe. Lower frequency corresponds to earlier time and vice versa. As an estimation, let us consider Model Ia in standard cosmology. From Eq.~(\ref{eq:gamma}), the DR particles with energy $\omega$ at present are created at a cosmic time $t_{e}$, satisfying $\omega a(t_{0})/a(t_{e})=m_{X}/2$. Due to entropy conservation, $g_{*S}(T)T^{3}a^{3}$ is a constant. Hence we can convert $t_{e}$ to temperature $T$ when DR is created,
\begin{equation}\label{eq:tem}
T=\frac{m_{X}}{2}\frac{g_{*S}^{1/3}(T_{0})}{g_{*S}^{1/3}(T)}\frac{T_{0}}{\omega}\, ,
\end{equation}
where $T_{0}=2.725\, \rm K$ is the present CMB temperature. DR frequency $10^{-3}\,{\rm eV}<\omega<10^{-2}\,{\rm eV}$ shown in Fig.~\ref{imgd} corresponds to an early universe temperature $80\,{\rm GeV}\lesssim T\lesssim 798\,{\rm GeV}$.

Finally we briefly discuss Model III and more general models. As an illustration, we use a Gaussian line profile instead of a $\delta$-function as used in Model Ia, 
\begin{equation}
\Gamma\left(\omega\frac{a(t_{0})}{a(t)},\rho_{X},T\right)= \frac{\Gamma_{X}\rho_{X}(t)}{\sqrt{\pi}\Delta}\exp\left[{-\frac{\left(\omega\frac{a(t_{0})}{a(t)}-\frac{m_{X}}{2}\right)^{2}}{\Delta^{2}}}\right]   \, , 
\end{equation}
where the parameter $\Delta$ characterizes the line width. We expect that as $\Delta$ increases to a large enough value, the kink caused by a PT will be smoothed out. In Model Ia, if we consider the thermal velocity of X, a Doppler broadening appears, which corresponds to a Gaussian line profile with $\Delta^{2}=m_{X}T_{X}/2$ for $T_{X}\ll m_{X}$, where $T_{X}$ is the temperature of X. If X is still in thermal equilibrium with Standard Model particles, $T_{X}$ is just the photon temperature $T_{\gamma}$. Otherwise, $T_{X}<T_{\gamma}$ when X becomes non-relativistic. For $m_{X}=20\,{\rm TeV}$ and $T_{X}\sim 100\,{\rm GeV}$, Doppler broadening only introduces a small correction to Fig.~\ref{imgd}. The DR spectrum in Model III is shown in the lower panel of Fig.~\ref{imga}. The total DR density is suppressed as before. If DR is created before PT, the peak frequency is redshifted, otherwise it is almost unchanged, as shown in the lower right panel of Fig.~\ref{imga}.

Using the same method we can generalize the result to the case of temperature-dependent intrinsic spectrum. There is one important exception---as discussed below, if the energy of DR when it is created is proportional to temperature $T$, we cannot learn about the universe evolution from the DR spectrum alone.

\begin{figure}[t]
  \centering
  \includegraphics[width=8.5cm]{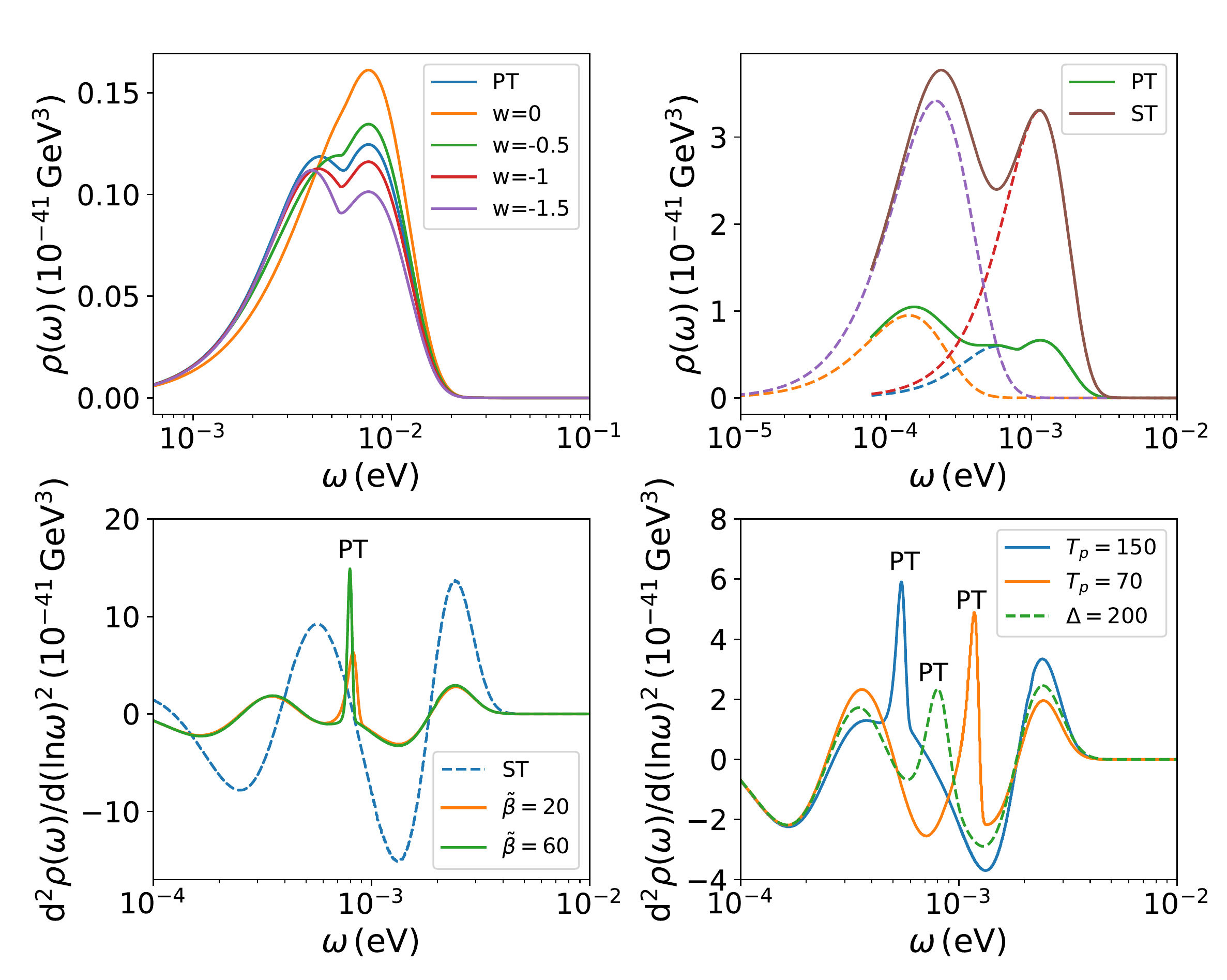}\\
  \caption{(\emph{Upper left}) DR spectrum in Model Ia with $m_{X}=20\,{\rm TeV}$ for PT and cosmology with a short period of early component N domination, whose equation of state is $p=w\rho$. We take $\alpha=3$, $\alpha_{\Lambda}=2.13$, and $T_{p}=100\,{\rm GeV}$ here. (\emph{Upper right}) The superposition of DR spectrum (solid lines) in Model Ia with $m_{X}=3\,{\rm TeV}$ and the thermal spectrum. The thermal spectrum is shown as two dashed lines on the left. (\emph{Lower left}) $\mathrm{d}^{2}\rho(\omega)/\mathrm{d}(\ln{\omega})^{2}$ shows a peak near PT. Here $\rho(\omega)$ includes the thermal spectrum and DR spectrum in Model Ia with $\alpha_{\Lambda}=3$, $T_{p}=100\,{\rm GeV}$. (\emph{Lower right}) The same as the figure in lower left with $\tilde{\beta}=20$, $\alpha_{\Lambda}=3$ but with different $T_{p}$ (solid lines). The dashed line is for Model III with $\Delta=200\,{\rm GeV}$ and $T_{p}=100\,{\rm GeV}$.}
  \label{imgf}
\end{figure}

\begin{figure}[t]
  \centering
  \includegraphics[width=8.5cm]{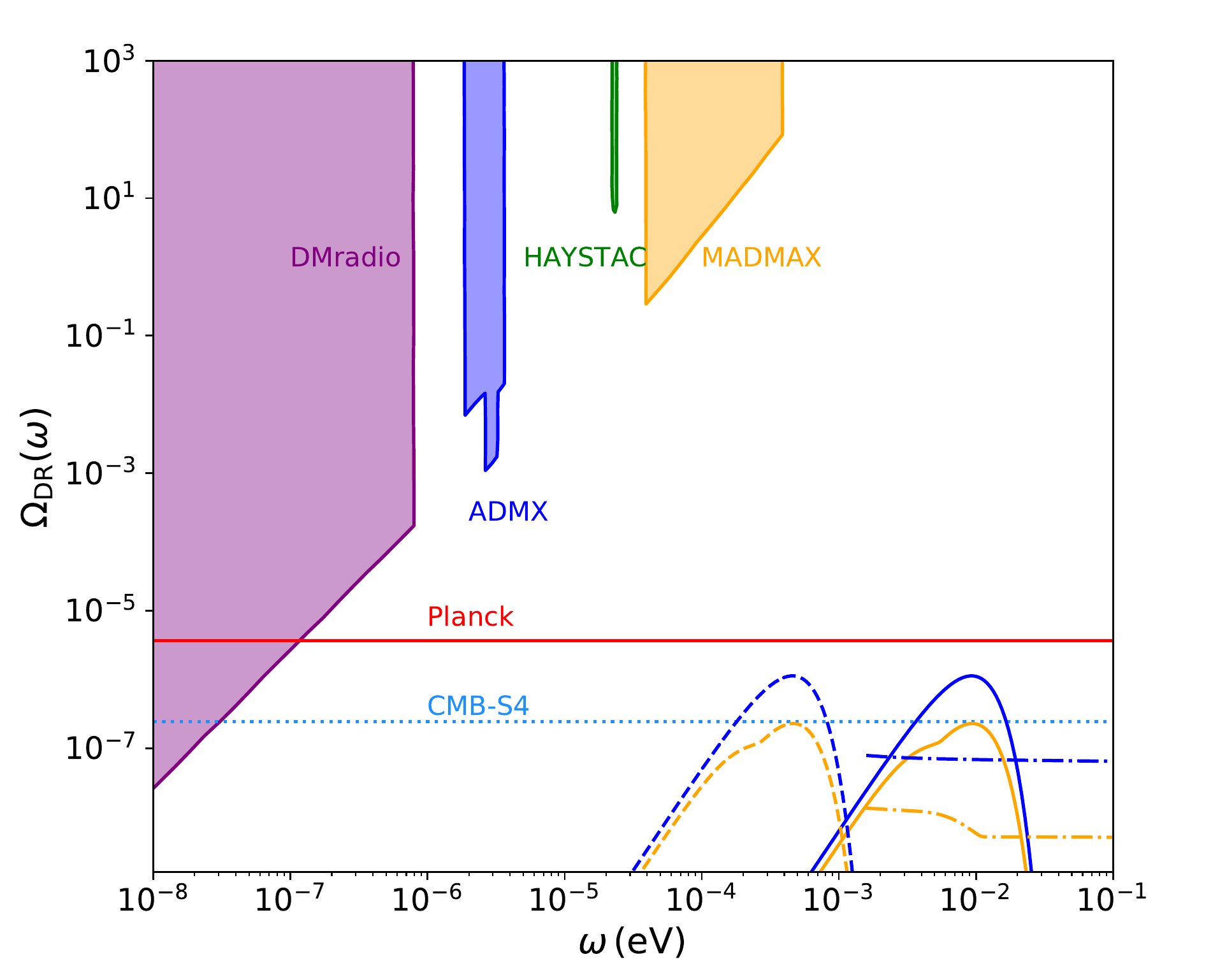}\\
  \caption{The CAB detection sensitivity~\cite{CaB} for an axion-photon coupling $g_{a\gamma\gamma}=0.66\times 10^{-10}\,{\rm GeV^{-1}}$. The sensitivity of Planck~\cite{planck} and projected sensitivity of CMB-S4~\cite{CMBS4} are plotted by assuming that the energy $\omega$ of CAB extends one order of magnitude.  We neglect axion mass here. We also plot the DR spectrum for Models Ia (\emph{solid}), Ib (\emph{dashed}), and II (\emph{dash-dot}) for both standard cosmology (\emph{blue}) and with PT (\emph{orange}). PT parameters are taken as $\alpha_{\Lambda}=3$, $T_{p}=100\,{\rm GeV}$, and $\tilde{\beta}=20$. We have taken $m_{X}=20\,{\rm TeV}$, $m_{Z}=19.5\,{\rm TeV}$, and $\Gamma_{X}=3\times 10^{-14}\,{\rm GeV}$ here. For Models Ia and Ib, initial densities are chosen such that DR contributes $\Delta N_{\rm eff}=0.1$ in the standard cosmology.}
  \label{imgx}
\end{figure}

The kink in DR spectrum in Models I and II is mainly due to the change of Hubble rate. Hence it is natural to ask whether other processes that modify the Hubble rate in the early universe produce similar effects. Consider a component N with an equation of state $p=w\rho$ with $w< 1/3$. The ratio of the density of N and radiation density reaches a maximum $\rho_{N}/\rho_{R}=\alpha$ ($\alpha\sim\mathcal{O}(1)$) at a temperature $T_{p}$ and then quickly decays to radiation with a constant decay rate $\Gamma_{N}$. The universe is still radiation dominated except near $T_{p}$. Such a process decreases the DR density and may produce a similar kink in DR spectrum. In the upper left panel of Fig.~\ref{imgf} we plot the DR spectrum for different $w$ with $\alpha=3$ and $\gamma=\Gamma_{N}/H(t_{p})=20$. $\alpha$ is related to reheating and can be roughly inferred from DR density observed and DR density in the standard cosmology. The case with $w=0$ is the early matter domination model, which produces a kink that is barely visible. Even if we increase $\alpha$, the kink does not become more apparent. A smaller $w$ produces a larger kink for a fixed $\alpha$. We also plot the DR spectrum for PT, with $\alpha_{\Lambda}=3\times0.71$ and $T_{p}=100\,{\rm GeV}$. The factor $0.71$ is introduced to give the same initial ECC density for the cases of $w=-1$ and PT. For $w=-1$, the shape of DR spectrum becomes the same as PT discussed before. From the depth of the kink we can determine whether $w=-1$, which is the case for PT. 


Besides the DR particles created from above models, there always exist a thermal population of DR particles. These DR particles should have a decoupling temperature $T_{d}$ higher than TeV scale, so that its interaction with standard model can be neglected. The temperature of the thermal spectrum is~\cite{CaB} $T=T_{0}(g_{*S}(T_{0})/g_{*S}(T_{d}))^{1/3}$. The thermal spectrum contributes $\Delta N_{\rm eff}=0.0268$. If a PT occurred after DR particle decoupling, the temperature of the thermal spectrum will be further suppressed, as $T'\approx T(1+\alpha_{\Lambda}/0.71)^{-1/4}$. The superposition of the thermal  and non-thermal spectra from Model Ia (neglect Doppler broadening) also exhibits a concave feature shown as the brown solid line in the upper right panel of Fig.~\ref{imgf}. However, as long as the PT is fast with $\tilde{\beta}\gtrsim 20$, the concave feature is different from the kink produced by PT. PT corresponds to a sharp turn in DR spectrum. If we have the ability to measure the DR spectrum precisely, the second derivative $\mathrm{d}^{2}\rho(\omega)/\mathrm{d}(\ln{\omega})^{2}$ can be obtained. The kink caused by PT corresponds to a distinct peak of $\mathrm{d}^{2}\rho(\omega)/\mathrm{d}(\ln{\omega})^{2}$ shown in the lower two panels in Fig.~\ref{imgf}. This is because $\mathrm{d}\ln{H}/\mathrm{d}\ln{t}$ changes quickly during a fast PT. The peak still exists if we instead consider Model III with a small width $\Delta$, shown as the dashed line in the lower right panel of Fig.~\ref{imgf}.

Consider the case where DR is made of axions. The sensitivity~\cite{CaB} of various experiments and the CAB from Models Ia, Ib, and II are shown in Fig.~\ref{imgx} in solid, dashed and dash-dot lines respectively. We do not include Model III in the figure because its spectra lie very close to Model Ia. We neglect the axion mass here. For the QCD axion, we may need to include the mass effect for both the spectrum and the detection sensitivity. We note that if we choose $m_{Z}$ to be very close to $m_{X}$, the whole CAB spectrum moves leftwards. If the axion mass is negligible, and there is a fine tuning between $m_{X}$ and $m_{Z}$ so that they are almost equal, it is possible that the CAB is within the sensitivity of the DMradio experiment.

\emph{Discussion.---} 
ECC naturally arises if a PT occurred in the early universe. If there are some DR particles created around the PT temperature, they will carry the information of the early universe. DR density will generally decrease if DR particles are created before or near a PT. The change in the shape of DR spectrum depends on DR production mechanisms. If the intrinsic spectrum is a $\delta$-function, like in our Models I and II, PT leaves a unique kink in the DR spectrum. For more general intrinsic spectra like in Model III, although the distortion may be smoothed out, we can still find the trace of PTs from a combination of peak frequency and DR density. In such case PT and early matter domination can only be distinguished with precise knowledge of DR production.

We note that if DR particles are created at temperature $T$ from relativistic particles, the DR particle energy when it is created is $\omega_{T}\sim T$. Because of the scale factor $a\sim 1/T$ (this is true if $g_{*S}$ can be treated as a constant), the energy of DR particle now is $\omega\sim\omega_{T}a/a(T_{0})$. We find that DR particles created at different temperatures turn out to have the same energy today, and information of the early universe evolution is degenerate. Hence for such processes, the DR spectrum alone is not a good clue for early universe evolution. We require that the DR production mechanism should involve at least one non-relativistic particle, in which case DR particles with different energy correspond to different emission time in the early universe.

DR density may be detected in CMB observations, like CMB-S4~\cite{CMBS4}, which is sensitive to $\Delta N_{\rm eff}\sim 0.02$. If we detect the DR and find its density smaller than expected, it is still not enough to reach the conclusion that an early-universe PT occurred. If we have the ability to measure the DR spectrum, the situation becomes much better. The spectrum provides information of the PT temperature and strength, which can be compared with the decrease of total DR density. If DR is made of axions, we may detect it in resonant cavity experiments that are used to search for axionic dark matter. These experiments are sensitive to the axion energy and can be used to detect the shape of CAB spectrum. However, for most hot axion production mechanisms, the current sensitivities are not enough for detection.

~\\
We thank Li-Xin Li for helpful discussions. ZW was supported by the National
Natural Science Foundation of China (Grant No.\ 11973014). LS was supported by
the National Natural Science Foundation of China (Grant Nos.\ 11975027 and
11991053), the National SKA Program of China (Grant No.\ 2020SKA0120300), and
the Max Planck Partner Group Program funded by the Max Planck Society.

\bibliography{refs}
\end{document}